\documentstyle[aps,prb,epsf,twocolumn]{revtex}
\include{psfig}
\begin{document}
\preprint{\footnotesize 
G\"oteborg Preprint APR 99-25, 26 June 1999, submitted to Phys. Rev. B}
\title{Internal and interfacial friction in the 
       dynamics of soft/solid interfaces}
\author{M. V. Voinova, M. Jonson and B. Kasemo}
\address{Department of Applied Physics, Chalmers University of Technology and 
G\"oteborg University,\\
SE-412 96, G\"oteborg, Sweden}
\maketitle
\begin{abstract}
We analyze theoretically the effect of friction on
quartz crystal microbalance (QCM) measurements that probe soft (viscoelastic)
films and biomolecular layers adsorbed from aqueous solutions. While
water provides a natural environment for biomolecules, an interface with
unknown rheological properties forms between the adsorbed soft molecular 
layer and the quartz substrate in the latter case. We investigate therefore 
the dynamics of soft films adsorbed onto a solid quartz surface within 
a continuum mechanics approach using both the Maxwell and the
Voight/Kelvin models of viscoelasticity and their combination. The rigorous 
expressions derived for the acoustic response of a quartz crystal oscillator,
accounting for both interfacial (sliding) friction and 
internal friction (viscosity), demonstrate that the QCM can be used as a
sensor for quantitative characteristization of friction effects as well as
for ``in situ'' measurements of mechanical properties of
 adsorbed biomolecular films. 
\end{abstract}
\section{Introduction}

A universal slip law for a moving interface between a solid
and a fluid experiencing shear stress, 
has been demostrated  recently \cite{1} for cases
when the no-slip condition breaks down. Experimentally, 
this situation can be addressed with a piezoelectric (quartz)
plate, which oscillates in thickness-shear mode while in
contact with a fluid medium \cite{2}. Traditionally, such quartz 
crystal oscillators have been utilized as a 
microbalance (QCM) for weighing a  negligible amount of  adsorbed mass 
in vacuum or in a gaseous environment \cite{3}. Current applications of 
the QCM in a vacuum or 
gas phase range from nanotribology experiments on thin films of
solidified gases \cite{4,5,6,7,8}
to  sensor measurements of viscoelastic properties of thin films 
deposited onto the quartz crystal surface \cite{7}. 
In recent liquid phase experiments \cite{11,12}, the
QCM technique has been adopted as a biosensor for measuring the 
properties of adsorbed polymer- and biomolecular films.
However, a finite amount of slippage arising from a weak coupling between
film and  oscillating substrate can change both the resonance frequency $f$
and the quality factor $Q$ of the oscillator and renormalize the results of 
measurements. Therefore, understanding the role of sliding friction 
in liquid QCM experiments is of key importance. 
An extensive review of the theory of sliding friction and the quartz crystal
microbalance method can be found in the recent book by Persson \cite{9}.
%

For a quantitive characterization of slippage effects
it is convenient to use the ratio between the shift $\Delta f$ of the
QCM resonance frequency and the inverse quality factor 
$ \Delta (Q^{-1})$. This measure was introduced in the QCM 
literature by Krim and Widom \cite{6} as a ``slip time'' $\tau$,
\begin{equation}
 \tau  = \frac {\Delta (Q^{-1})}{4 \pi|\Delta f |} \, .
\label{eq.one}            
\end{equation}
In the following we prefer to refer to the dissipation factor 
$D\equiv Q^{-1}$ rather than to the inverse quality factor of the 
oscillator \cite{11}. The shifts $\Delta D$ and $\Delta f$ can be
measured simultaneously in QCM experiments; one always finds
$\Delta D$ to be positive while the
resonance frequency shift $\Delta f$ is negative since mass loading 
lowers the oscillator resonance frequency 
compared to the unloaded state in vacuum.

According to the results of Krim's group \cite{8}, a
partial decoupling of the overlayer should occur when $2\pi f \tau\geq 0.5$.
Typically, a quartz crystal oscillates with frequency $f \sim 10$~MHz
and therefore decoupling starts when $\tau\geq\tau_{c}\simeq 10^{-8}$~s. 
For instance, in QCM experiments with thin water films \cite{11}
 $\Delta f \sim 1-10$~Hz, $\Delta D \sim 10^{-7}-10^{-6}$ 
and hence $\tau \sim 10^{-9}-10^{-7}$~s,  which is 
quite close to $\tau_{c}$. It follows that a relatively low decoupling 
threshold can lead to noticeable interfacial friction effects even in 
the case of molecularly thin films. 

In the nanotribology QCM-experiments of Krim and collaborators, 
where they 
studied films of solidified gases, condensed onto a solid substrate
\cite{4,5,7,8}, 
the interfacial friction coefficient $\gamma_f$ for a thin 
rigid film was found to be inversely proportional to the slip time and 
proportional to the surface mass density $m_{f}=\rho_{f} h_{f}$ of
the film,
\begin{equation}
 \gamma_{f} = m_f/\tau. 
\label{eq.two}                 
\end{equation}
This result can also be viewed as the definition of the slip time
$\tau$,  which is \cite{8}
the characteristic time it takes for the film
velocity to decrease by a factor $1/e$.
However, for soft interfaces Eq.~(\ref{eq.two}) becomes invalid 
because of additional viscous dissipation of energy in the material.
 
In this article we analyze theoretically the QCM response, accounting 
for both viscosity and  interfacial friction effects, and derive how 
the sliding friction coefficient is related to the values of 
$\Delta f$ and $\Delta D$, which can be measured. Throughout the paper
we use the term ``soft''  to characterize properties which are opposite
to ``rigid", i.e., the viscoelasticity of the material.

We will show that under certain conditions, the dissipation factor
$\Delta D$ can have a maximum as a function of the viscosity as well 
as of the sliding friction of the QCM-probed material. We derive the 
exact expressions for the maximal dissipation shift.
The general analysis presented here is valid for arbitrary layer 
thickness. Sliding friction  of ultrathin films, adsorbed from a
gaseous (or liquid) phase, emerges as a special limiting case. 

From a practical point of view, one of the particularly important 
topics treated here is concerned with liquid- and biosensor QCM 
applications. Here the oscillator, with its adsorbed biomolecular layer,
operates in a bulk viscous liquid (water), which provides a natural 
environment for {\em in situ} measurements. We have modeled this 
system as a layered soft (viscoelastic) material 
covered by a bulk newtonian liquid.  Linear viscoelasticity of 
different biological interfaces has been analysed for
both Maxwell and Kelvin/Voight schemes and their combination.

\begin{figure}
\centerline{\psfig{figure=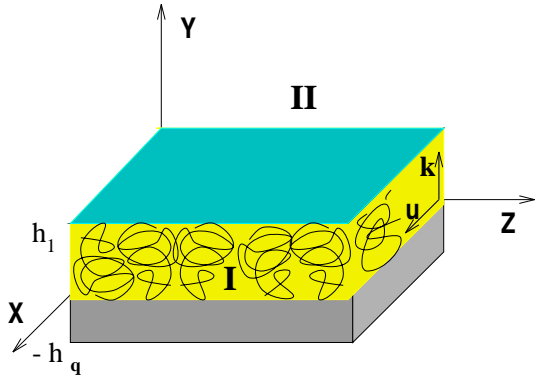,width=7cm}}
\vspace*{7mm}
\caption{Schematic depiction of a quartz crystal oscillator
(microbalance) loaded on one side by a
layered medium. Layer I could, {\em e.g.}, be a thin biomolecular
layer, layer II could be, e.g., bulk water.}
\end{figure}

\section{Damping of a quartz oscillator by slipping soft material.}

Let us consider a quartz crystal microbalance (QCM), oscillating in 
thickness-shear mode and on one side covered by a soft adsorbed layer 
(as depicted schematically in Fig.~1). We describe the motion of the QCM 
as the forced vibration of a damped oscillator driven by periodic 
force \cite{6,11}.

When mass is deposited onto the crystal surface,
the resonance frequency changes. For a small amount of mass 
or a molecularly thin film rigidly attached
to the QCM surface, the shift of resonance frequency $\Delta f$ is
 proportional to the density $\rho$ and thickness $h$ 
of the overlayer \cite{3},
 while there is no dissipation shift $\Delta D$.
When the oscillator operates in a bulk liquid environment, both  
$\Delta f$ and $\Delta D$ are finite and 
depend on the density $\rho$ and the
viscosity 
$\eta$ of the liquid.
Classic results are Kanazawa and Gordon's resonance frequency
shift of a QCM immersed in a bulk liquid \cite{15}:
$$
\Delta f \approx - \frac{1}{2\pi m_q}
\sqrt{\frac{\rho\eta\omega}{2}} \, ,
$$ 
and Stockbridge's result for the QCM dissipation shift \cite{16}:
$$
\Delta D\approx \frac{1}{m_q}\sqrt{\frac{2\rho\eta}{\omega}} \, .
$$ 
Here $m_q = \rho_q\, h_q$ is the surface mass density of the quartz plate,
$\rho_q$ and $h_q$ is the mass density and thickness of the quartz, 
respectively.
In the general case of a layered viscous (viscoelastic) medium, both  
$\Delta f$ and $\Delta D$ are complex functions of thickness, mass and 
viscoelastic properties of the material \cite{13,17,18,19}. 
These characteristics 
can be derived from the acoustic impedance $Z(\omega)$ of the system.

The acoustic impedance
\begin{equation}
Z=Z(\omega)=Z^{'}-iZ^{''}
\label{eq.three}
\end{equation}
is equal to the mechanical impedance
 per unit area. The real part $Z^{'}$ of $Z(\omega)$ is a resistive term, 
 which is proportional to the average energy dissipation per unit time, 
while the imaginary part $Z^{''}$ is a  reactive term associated with 
the inertia of the oscillator \cite{6,7}.
For a material rigidly attached to the quartz surface, the so-called 
no-slip boundary conditions are valid, i.e. the velocity of the
quartz surface is equal to the velocity of the adjacent layer. 
 Below we indicate acoustic characteristics determined in the absence of
 slippage (no-slip boundary conditions) by a tilde sign ( $\tilde{ }$ ) over
the appropriate symbol. 
The shifts in the
QCM resonance frequency and in the dissipation factor in the absence of 
slippage are given by the expressions \cite{6}
\begin{equation}
\Delta\tilde{f} = -{\rm Im}\left( \frac{\tilde{Z}}{2\pi m_{q}} \right)
=\frac{\tilde{Z^{''}}}{2\pi m_{q}}
\label{eq.four} 
\end{equation}
and
\begin{equation}
\Delta\tilde{D} = {\rm Re}\left( \frac{\tilde{Z}}
{\pi f m_{q}} \right) =\frac{\tilde{Z^{'}}}{\pi f m_{q}}\, .
\label{eq.five}
\end{equation}

Let us now consider the effect of potential slip arising from a weak 
coupling of film and oscillating substrate. In this case the no-slip 
boundary condition breaks down and a finite difference between the 
velocity $\dot{q_0}$ of the quartz surface and the velocity $v_{0}$ 
of the adjacent layer at the interface (index 0) appears leading to
an interfacial friction force, $F_s$, where
\begin{equation}
F_{s}= \gamma_{s} (\dot{q_0} - v_{0}).
\label{eq.six}
\end{equation}
Here $\gamma_{s}$ is the proportionality 
coefficient characterizing sliding friction. It is a constant
in the low shear rate approximation. 
Slippage can also be characterized by a slippage coefficient
 $\lambda$ which is simply the inverse of the sliding friction 
coefficient \cite{14},
\begin{equation}
\lambda = \gamma_{s}^{-1}.
\label{eq.seven}
\end{equation}
The no-slip condition corresponds to $\lambda \to 0$, 
while the opposite limit $\lambda \to \infty$ gives infinite slippage 
and a complete decoupling of the overlayer from the substrate. 
Effects of slippage  can easily be included in the equation of
motion for the oscillator. Following the damped oscillator model 
of Ref.~\cite{11}, we may express the shift of resonance frequency and 
dissipation factor in the presence of slippage as 
\begin{equation}
\Delta f \approx -Im \left( \frac{\tilde{Z}}{2\pi m_{q}(1 + 
\lambda \tilde{Z})}\right)
\label{eq.eight}
\end{equation}
\begin{equation}
\Delta D \approx  Re \left( \frac{\tilde{Z}}{\pi f m_{q}(1 + 
\lambda \tilde{Z})}\right),
\label{eq.nine}
\end{equation}
where as already mentioned $\lambda=1/\gamma_s$.

Following the approach of Krim's group \cite{4,5,6,7}, we introduce the
 characteristic slip time $\tau_{s}$ as the ratio between
the shift of the dissipation factor and the change in resonance 
frequency change,
\begin{equation}
\tau_{s}= - \frac{\Delta D}{4\pi \Delta f},
\label{eq.eleven}
\end{equation}

Proceeding by means of an analogy, we define another ratio
with dimension of time,
\begin{equation}
\tau_{0}= - \frac{\Delta\tilde{D}}{4\pi \Delta\tilde{f}},
\label{eq.twelve}
\end{equation}
in order to characterize the temporal QCM response when no-slip 
boundary conditions apply.
Defined in this way, the characteristic time $\tau_{0}$ can be 
attributed solely to the effects of internal friction (viscosity) 
on the damping of the QCM, i.e,  $\Delta\tilde{D}$ is
proportional to the resistive term (\ref{eq.five}) in the 
acoustic ``no-slip'' impedance $\tilde{Z}$.

Using Eqs.~(\ref{eq.eight})-(\ref{eq.twelve}), we may express the 
slippage coefficient $\lambda$ and the interfacial friction coefficient 
$\gamma_{s}$ as functions of $\tau_{s}$ and $\tau_{0}$. One finds
\begin{equation}
\lambda= \frac{\omega \tau_{0} - 
\omega \tau_{s}}{2\pi \Delta\tilde{f} m_{q}} \, ,
\label{eq.thirteen}
\end{equation}
and
\begin{equation}
\gamma_{s}= 2\pi \Delta\tilde{f} m_{q}
\frac{1}{\omega \tau_{0} - \omega \tau_{s}} \, .
\label{eq.fourteen}
\end{equation}

Let us consider a viscoelastic medium of finite thickness $h$ in
contact with the quartz oscillator (Fig.~1). It is known, that
the acoustic response of the oscillator 
strongly depends on the overlayer thickness. ``Thin'' or ``thick'' films
refers to the film thickness being
smaller or greater than the inverse values of the decay constant $\alpha$ 
and propagation constant $k$ of the acoustic waves propagating 
in the coated quartz plate \cite{18}.

In the limiting case of an ultrathin film ($h \alpha \ll 1, h k \ll 1$),
acoustic shear waves can propagate through the adsorbed film without 
dissipation. Such a layer has a Sauerbrey solid-like response 
\cite{3}, for which
$$
\Delta \tilde{f} = - f_{0}\frac{m_{f}}{m_{q}}\, ,
\quad \Delta \tilde{D}=0 \, ,
$$
and where $m_{f}=\rho\, h_{f}$ is the mass per unit area of the film.
Substitution of $\Delta \tilde{f}$ and $\Delta \tilde{D}$ in
Eq.~(\ref{eq.fourteen}) gives the coefficient of interfacial friction.
For a molecularly thin overlayer one finds
$$
\gamma_{s0}= m_{f}/\tau_{s}, 
$$
which reproduces the result of Krim and Widom \cite{6}.
For slipping soft material of arbitrary thickness,
one must take into account both interfacial friction and
 viscosity and use Eqs.~(\ref{eq.eight})-(\ref{eq.fourteen}).  

An analysis of the general expressions (\ref{eq.four}), (\ref{eq.five}) 
and (\ref{eq.nine}) allows the dissipation factor $\Delta D(\lambda)$ 
to be expressed as a function of slippage as
\begin{equation}
\Delta D(\lambda)=  \frac{1}{\pi f m_{q}}\frac{\tilde{Z^{'}}+ 
\lambda ({\tilde{Z^{'}}}^2+{\tilde{Z^{''}}}^2)}
{1+\lambda^2({\tilde{Z^{'}}}^2 + {\tilde{Z^{''}}}^2) +
2\lambda\tilde Z^{'}}\, .
\label{eq.fifteen}
\end{equation}
From Eq.~(\ref{eq.fifteen}) it follows that the dissipation shift
is a nonmonotonic function of slippage $\lambda$, which
peaks for a critical value $\lambda^{\ast}$. 
If we take the limit  $\tilde{Z^{'}}=0$ corresponding to a rigid thin 
film, the maximal dissipation is
\begin{equation}
\Delta D(\lambda^{\ast})= - \frac{\Delta\tilde{f}}{f}\, .
\label{eq.sixteen}
\end{equation}
The dissipation is maximal in this case ($\omega \tau_{0}= 0$)
 when the condition
$$
\omega \tau_{s}=1 
$$
is satisfied, a criterion for sliding thin 
rigid films first found by Krim and Widom \cite{6}.

The condition (\ref{eq.sixteen}) is readily understandable. 
For a non-dissipative medium,
the damping of the oscillator is proportional to the inertial
contribution $\Delta\tilde{f}/f \sim m_{f}$ associated with 
the slippage of the added mass
pushed along the quartz surface. A rigid thin film will give rise to 
maximal dissipation when the sliding friction reaches
the value 
\begin{equation}
\gamma^{\ast}=m_{f}\cdot \omega \, , 
\label{eq.seventeen}
\end{equation}
or equivalently when the slippage coefficient is
\begin{equation}
\lambda^{\ast}= 1/(m_{f}\cdot \omega) \, .
\label{eq.eighteen}
\end{equation}
This maximum value of the dissipation factor is 
\begin{equation}
\Delta D(\lambda^{\ast})=m_{f}/m_{q}.
\label{eq.nineteen}
\end{equation}
The lower the surface mass density is, the larger is the slippage
coefficient $\lambda^{\ast}$ for which the dissipation factor is
maximal, and the smaller is this maximal value (Fig.~2). An important 
consequence of the result~(\ref{eq.nineteen}) is that 
the maximum value of the dissipation factor 
is frequency independent. Therefore, one may consider the presence of
such a frequency independent maximum as a test of whether slippage occurs in
dynamic QCM measurements; if the dissipation peak remains constant when the
oscillation frequency $\omega$ is varied, the peak can be attributed to 
sliding friction.

\begin{figure}
\centerline{\psfig{figure=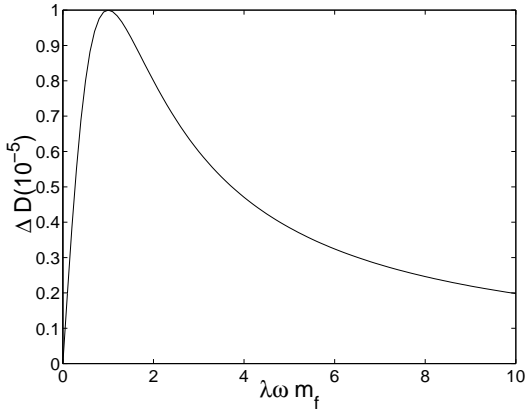,width=7cm}}
\vspace*{7mm}
\caption{Sketch of typical nonmonotonic dependence of the
dissipation factor $\Delta D$ on the dimensionless slippage coefficient 
$\lambda\, \omega \, m_f$ for a thin, rigid film adsorbed onto the 
quartz plate. The calculation was done for film thickness $h_f=1 \mu$m
and film mass density $\rho_f = 1$~g/cm$^3$.
}
\end{figure}

Experimentally, the sliding friction is an external parameter which
can be changed, e.g., by varying pressure or temperature.
Maxima in the dissipation has been observed experimentally 
\cite{4,5} for solid
molecularly thin films sliding along a quartz surface 
oscillating in gaseous environments. The maxima appeared when the 
the condition $\omega\tau_{s}=1$ was met. In these experiments, the effect of
interfacial friction on the QCM damping was obvious from the fact
that viscous losses are negligible both in a gas and in rigid molecularly 
thin films. 
However, in experiments with soft films, 
the dissipation factor can reach a maximum as temperature
 or pressure is varied, since these affect the internal friction
(shear viscosity) of the material. When slippage is absent (or 
negligibly small), viscoelastic effects dominate and the dissipation 
maximum becomes frequency dependent (see Section 3C).

In case of sliding thin viscoelastic overlayers, internal friction 
(viscosity) leads to additional but small contributions to the QCM
damping since $\Delta\tilde{D} \neq 0$. As a result the position
of the dissipation maximum is only slightly shifted.
On the contrary, the viscosity of soft materials of
finite thickness will significantly influence the damping and 
produce a ``viscous''-type dissipation peak.
In order to separate these effects, we study below the no-slip limiting
case --- when the interfacial slippage vanishes --- for the viscoelastic 
film dynamics .

\section{Acoustic impedance of a QCM loaded by a viscoelastic layer}

Viscoelasticity is a common feature of complex fluids and in polymer
rheology. A limiting case is that of highly viscous fluids - amorphous solids 
and glasses. Polymers exhibit linear viscoelasticity if both deformations
and deformation rates are small 
\cite{21,22}. In the linear viscoelastic regime,
there is proportionality between stored energy and strain and between 
strain rate and dissipated energy \cite{22}. The most well known 
mathematical formulations of linear viscoelasticity are the Maxwell 
model of relaxation in highly viscous fluids and the Kelvin/Voight
model of viscoelastic solids.

\subsection{Highly viscous Maxwell fluids}

Let us derive the acoustic impedance of a QCM covered by an overlayer of
highly viscous material. 
A very viscous complex fluid responds initially to deformation in the same
way as an elastic solid, but after the deformation stops the fluid relaxes. 
This relaxation processes can be characterized by the viscous relaxation time
$\tau_{M}$, which is a measure of the time it takes for the remaining  stress
to be damped \cite{21}. Such viscoelastic behavior
 of the complex fluid
can be characterized by the shear viscosity coefficient $\eta$ and the 
shear elasticity modulus $\mu$. The characteristic 
relaxation time $\tau_{M}$ is of the
order of their ratio \cite{21}:
$$
\tau_{M} \sim \eta/\mu.
$$

Viscoelastic properties of a complex fluid
can be treated within the Maxwell model which represents a mechanical
analogue of viscoelasticity via a dashpot (viscous newtonian element)
and a spring (elastic hookian element), arranged in series (Fig.~3a).
A Maxwell fluid subject to quartz crystal oscillations of frequency
$\omega$ behaves as a newtonian viscous fluid with 
shear viscosity $\eta$ if $\tau_{M} \ll \omega^{-1}$.
In this limit, the relation between the stress tensor $\sigma_{ik}$ and the 
strain tensor $u_{ik}$ is the same as for the ordinary 
newtonian fluid with $\sigma_{ik} = 2i\omega\eta u_{ik}$.
In the opposite case of large $\omega \gg \tau^{-1}$ , the maxwellian 
fluid responds to shear stress as a solid. The strain-stress relation 
$\sigma_{ik} = 2\mu u_{ik}$ describes the elasticity of a solid body 
\cite{21}. 

\begin{figure}
\centerline{\psfig{figure=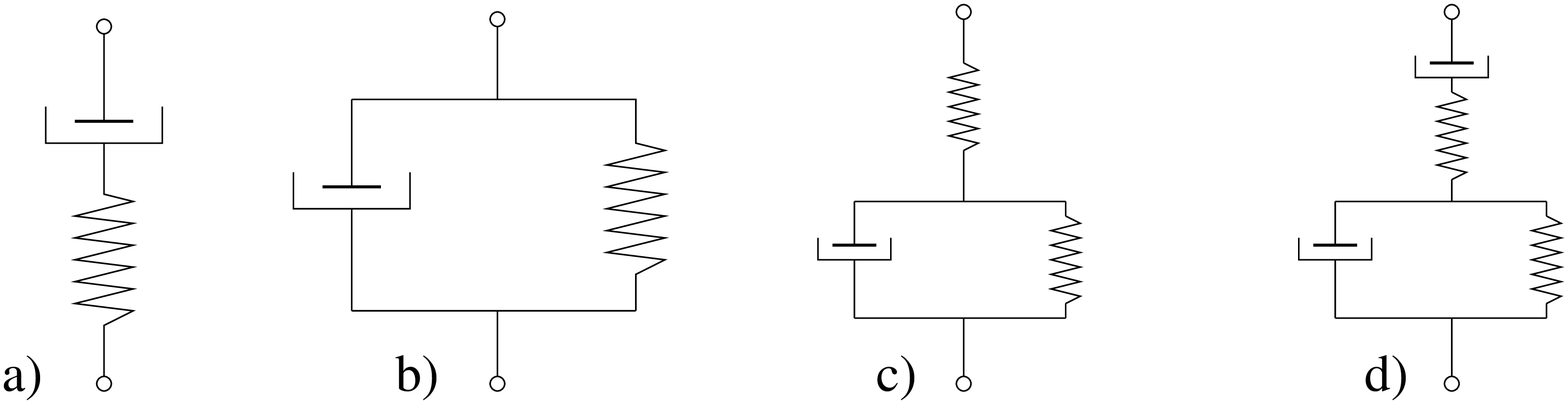,width=7cm}}
\vspace*{7mm}
\caption{Basic models for the linear viscoelasticity of 
different materials. a) Maxwell fluid; b) Voight/Kelvin 
viscoelastic solid; c)  and d) more complex rheological models of
viscoelastic solids.}
\end{figure}

The equation of motion for a maxwellian fluid,  which includes 
both the limits of slow- and  fast motion, is given by the expression 
\begin{equation}
2\mu\frac{d u_{ik}}{dt} = \frac{d\sigma_{ik}}{dt}+
\frac{\sigma_{ik}}{\tau_{M}}  \, .
\label{eq.twenty}
\end{equation}
Equation (\ref{eq.twenty})  assumes that the internal stress is 
exponentially damped on the characteristic time scale 
$\tau_{M}$  after the applied
stress is removed \cite{21}. 
Using Eq.~(\ref{eq.twenty}) we find the stress-strain
relation for maxwellian fluid to be of the form
\begin{equation}
\sigma_{ik} = \frac{2\mu u_{ik}}{1-i/\omega\tau_{M}} \, , 
\label{eq.twentyone}
\end{equation}
which can be rewritten as
\begin{equation}
\sigma_{ik} = 2\mu^{\ast} u_{ik}\, . 
\label{eq.twentytwo}
\end{equation}
Here $\mu^{\ast}_{M} \equiv \mu^{'}_{M}+i\mu{''}_{M}$
 is the complex shear modulus of a Maxwell fluid. The real part (storage
modulus) is
\begin{equation}
\mu^{'}_{M}= \frac{\eta ^{2}\omega ^{2}\mu}{\mu^{2}+\omega^{2}\eta^{2}}\, , 
\label{eq.twentythree}
\end{equation}
while the imaginary part (loss modulus) is
\begin{equation}
\mu{''}_{M} = 
\frac{\omega\eta \mu^{2}}{\mu^{2}+\omega^{2}\eta^{2}} \, .
\label{eq.twentyfour}
\end{equation}

Using the strain-stress relation (\ref{eq.twentyone}) and the conservation of
momentum density,  
$$
-\frac{\partial \Pi_{ik}}{\partial x_{k}}= \frac{\partial}{\partial t}
(\rho v_{i}) \, ,
$$
we get a wave equation, 
\begin{equation}
\mu^{\ast}\frac{\partial^2 u_x(y,t)}{\partial y^2} = 
-\rho \omega^2 u_x (y,t) \, ,
\label{eq.twentyfive}
\end{equation}
for bulk shear waves propagating in the system. Here $u_x(y,t)$ is the
a component of the displacement vector (Fig.~1). Solving equation 
(\ref{eq.twentyfive}) with
no-slip boundary conditions, we find \cite{26}
 the acoustic impedance of the  
maxwellian viscoelastic fluid overlayer of thickness $h$
to be given by the expression
\begin{equation}
\tilde Z_{h} = \frac{\xi \mu^{\ast}}{i\omega} \frac{1-e^{2\xi h}}
{1+e^{2\xi h}} \, . 
\label{eq.twentysix}
\end{equation}
Here $\xi = \alpha + ik$; $\alpha$ is the decay constant and $k$ is the
propagation constant. These quantities are given by formulae
\begin{equation}
\alpha _{M}= \frac{1}{\delta}\sqrt{\sqrt{1 + \chi^{2}_{M}} - \chi_{M}}
\label{eq.twentyseven}
\end{equation}
\begin{equation}
k_{M}= \frac{1}{\delta}\sqrt{\sqrt{1 + \chi^{2}_{M}} + \chi_{M}} \, .
\label{eq.twentyeight}
\end{equation}
In Eqs.~(\ref{eq.twentyseven}) and (\ref{eq.twentyeight}) the viscous 
wave penetration depth,
$\delta =(2\eta/\rho\omega)^{1/2}$ is \cite{20} the distance over 
which the  amplitude falls off by a factor of $e$, 
the viscoelastic ratio,  $\chi = \mu^{'}/\mu^{''}$, is the ratio
between real part (storage modulus) and the imaginary part (loss modulus)
of the complex shear modulus. For Maxwell material
$$
\chi_{M}=\eta\omega/\mu .
$$
Below we will characterize the material by the viscoelastic 
ratio $\chi$ rather then by the relaxation time $\tau_{M}$. These
quantities are 
related, since $\chi_{M} \approx \omega\tau_{M}$.  
 
\subsection{Thin Maxwell-fluid overlayers}

Here we derive the shifts $\Delta f$ and $\Delta D$ characteristic of a
 Maxwell fluid thin film rigidly
attached to the surface of a QCM oscillating in vacuum.
By a series expansion in $h\alpha\ll 1$ and $h k \ll 1$, which is permissible
in the special case of a thin overlayer, we find from 
(\ref{eq.twentysix}) - (\ref{eq.twentyeight}) and from
(\ref{eq.four}) and (\ref{eq.five}) that
\begin{eqnarray}
\Delta \tilde{f} &\approx& -\frac{h\rho\omega}{2\pi m_q} \left(1
 + \frac{h^2\rho\omega^2}{3\mu}\right)
\label{eq.twentynine}\\
\Delta \tilde{D}&\approx& (2h^3\rho^2\omega)/(3 m_{q}\eta)
\label{eq.thirty}
\end{eqnarray}
It should be noted, that Eq.~(\ref{eq.thirty}) has the same form as 
the dissipation factor
for a thin pure viscous layer with constant shear viscosity 
coefficient $\eta$.

From expression (\ref{eq.twentynine}) we obtain for the 
equivalent mass $m_M$ of the film 
\begin{equation}
m_{M} = m_{f}(1+ h^2\rho\omega^2/3\mu)
\label{eq.thirtyone} 
\end{equation}
which differs from the actual mass $m_{f}$ and depends on the 
shear elasticity $\mu$ of the material.
Our results (\ref{eq.twentynine})-(\ref{eq.thirtyone}) are in agreement with
those of Johannsmann {\em et al.} \cite{10}, who calculated  
and measured 
%
%
by QCM the equivalent mass of a viscoelastic Langmuir-Blodgett film.
The characteristic ``viscous'' time $\tau_0$ for a thin Maxwell film
is, therefore
\begin{equation}
\tau_{0} \approx (h^2 \rho)/(3\eta) \, . 
\label{eq.thirtytwo}
\end{equation}
Because $\tau_0$ is defined as the ratio between the experimentally measured
values of $\Delta \tilde{D}$ and
$\Delta \tilde{f}$, one can calculate from Eq.~(\ref{eq.thirtytwo}) the
viscosity of a thin Maxwell film  if the film mass density and thickness are
known or measured separately (e.g., by neutron scattering  or ellipsometry
methods). 

Let us now proceed with a comparison between the results for a Maxwell fluid
with those for a thin viscoelastic solid film.
  
\subsection{Solid viscoelastic Voight/Kelvin overlayers}

In the Voight/Kelvin scheme (Fig.~3b) for describing visoelastic solid
materials, a complex viscoelastic shear modulus, $\mu^{\ast}_{V} \equiv
\mu^{'}_{V}+i\mu{''}_{V}$, is used. Its real part is the storage
modulus 
\begin{equation}
\mu^{'}_{V} = \mu \, ,
\label{eq.thirtythree}
\end{equation}
while the imaginary part is the loss modulus
\begin{equation}
\mu{''}_{V} = \omega \eta \, .
\label{eq.thirtyfour}
\end{equation}

The dynamics of viscoelastic solids can also be characterized 
by the retardation time $\tau_{V} \sim \eta_{V}/\mu_{V}$ \cite{23} or
by the viscoelastic ratio 
\begin{equation}
\chi_{V} = \frac{\mu}{\eta\omega} \, .
\label{eq.thirtyfive}
\end{equation}

The acoustic impedance of a quartz oscillator in the absence of slip
is given by expression (\ref{eq.twentysix}) 
where $\xi_{V}=\alpha_{V}+ik_{V}$ and
\begin{eqnarray}
\alpha_{V} &=& \frac{1}{\delta}\sqrt{\frac{\sqrt{1 + \chi_{V}^2} - 
\chi_{V}}{1 + \chi_{V}^2}}
\label{eq.thirtysix} \\
k_{V}&=& \frac{1}{\delta}\sqrt{\frac{\sqrt{1 + \chi_{V}^2} + 
\chi_{V}}{1 + \chi_{V}^2}} \, , 
 \, \;
\delta=\sqrt{\frac{2\eta}{\rho\omega}} \, .
\label{eq.thirtyseven}
\end{eqnarray}
In the same manner as in the previous section,
 we find  from (\ref{eq.twentysix}) taken together with 
(\ref{eq.thirtysix}) and (\ref{eq.thirtyseven}) that
\begin{equation}
\Delta \tilde{f}  \approx -\frac{1}{2\pi m_{q}}h\rho\omega \left(1
 + \frac{2h^2 \chi_{V}}{3\delta^2 (1+\chi_{V}^2)}\right)\\
\label{eq.thirtyeight}
\end{equation}
\begin{equation}
\Delta \tilde{D} \approx \frac{2h^3\rho\omega }{3\pi f  m_{q}} 
\frac{1}{\delta^2(1+\chi_{V}^2)}, ~
 \delta =\sqrt{\frac{2\eta}{\rho\omega}} \, .
\label{eq.thirtynine}
\end{equation}

The characteristic time $\tau_{0}$ depends on  frequency and 
viscoelastic moduli of the material. One finds   
$$
\tau_{0} \approx \frac{h^2 \rho\eta\omega^2}{3(\mu^2 +\eta^2\omega^2)} \, .
$$
We therefore obtain the equivalent mass of the Voight/Kelvin layer as
\begin{equation}
m_{V} = m_{f}\left(1+ \frac{h^2 \mu\omega^2\rho}{3(\mu^2 +
\eta^2\omega^2)}\right) \, ,
\label{eq.forty}
\end{equation}
which includes a correction of the true film mass density
 $m_{f}$ due to the viscoelasticity of the material.

Equations~(\ref{eq.thirtynine}) and (\ref{eq.forty}) together with $\tau_{0}$
allow us to calculate the viscoelastic parameters $\mu_{V}$ and $\eta_{V}$
from simultaneously measured values of $\Delta \tilde{f}$ and 
$\Delta \tilde{D}$ if it possible to control the film thickness $h$
independently.

From expression (\ref{eq.thirtynine}) it follows that if during the
experiment the shear elasticity is constant but the viscosity varies, the
dissipation factor as a function of shear viscosity $\eta$ will have a maximum
when $\eta^{\ast}=\mu/\omega$ (or $\chi_{V}=1$, respectively) (Fig.~4). 
And vice versa, in experiments with constant viscosity but varying shear
elasticity $\mu$,  the resonance frequency shift will have a maximum for
$\mu^{\ast}=\eta\omega$ ($\chi_{V}=1$) (Fig.~5). 
The maximum value of the dissipation factor 
$$
\Delta \tilde{D}(\eta^{\ast})= \frac{h^3\rho^2\omega^2}{3 m_{q}\mu}
$$
depends on the vibration frequency $\omega$ squared. 
Since the dissipation peak due to sliding friction was found to be frequency
independent, this strong frequency dependence of
 $\Delta \tilde{D}(\eta^{\ast})$ can be used as a test to see if slippage
is absent in QCM ``viscoelastic'' thin film measurements. 
Note that the ``viscous'' peak value $\Delta \tilde{D}(\eta^{\ast})$
is weaker by a factor $h^2\rho\omega^2/\mu$ than the 
peak value $\Delta D(\lambda^{\ast})$ of Eq.~(\ref{eq.nineteen}). 

It should be noted, that the frequency
behavior of the QCM characteristics in the linear viscoelasticity region,
{\em viz.} $\Delta f(\omega)$,  can also test for the type of material -
either viscoelastic fluid or viscoelastic solid. This is because 
from our results (\ref{eq.twentynine}) and (\ref{eq.thirtynine}) for 
$\Delta f_{M}$ and $\Delta f_{V}$, it follows that we expect 
a straight line  if $\Delta f(\omega)$ is plotted vs. $\omega^2$ for the
case of a Maxwell fluid and a deviation from a straight line for other types of
viscoelastic materials. In the case of Maxwell fluid, the slope of the straight
line gives the elastic modulus $\mu$ of the material. (The viscosity $\eta_{M}$
of the Maxwell fluid film can then be deduced directly from the measured
values of $\Delta D$ or $\tau_{0}$ if the film thickness is known.)

\begin{figure}
\centerline{\psfig{figure=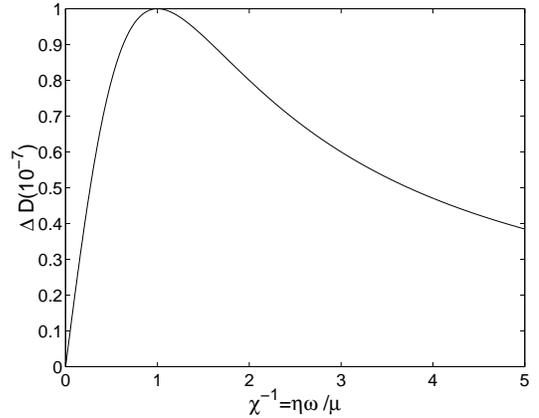,width=7cm}}
\vspace*{7mm}
\caption{ Calculated dissipation factor $\Delta D$
of a quartz crystal microbalance
(QCM) covered with a thin film of a
Voight/Kelvin viscoelastic solid as a function of its inverse viscoelastic
 ratio $\chi^{-1}=\eta\omega/\mu$; $h_1 = 1$~$\mu$m, 
$\omega=2\pi \cdot 10$~MHz,
$\mu=10^{10}~$dyn/cm$^2$, $\rho = 1$~g/cm$^3$.}
\end{figure}
\begin{figure}
\centerline{\psfig{figure=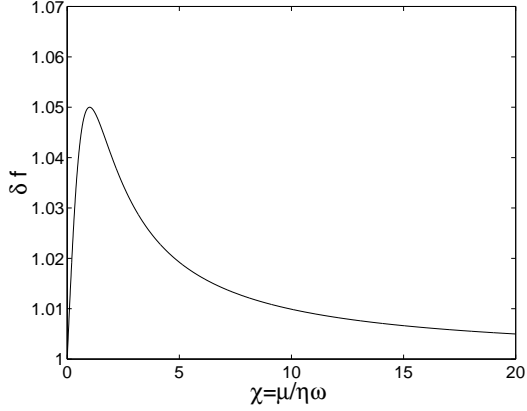,width=7cm}}
\vspace*{7mm}
\caption{ Relative resonance frequency shift $\delta f$ of a QCM
due to an adsorbed viscoelastic solid thin film; $h_1 = 0.1$~$\mu$m, 
$\omega=2\pi \cdot 10$MHz, $\eta= 1$~dyn$\cdot$c/cm$^2$, 
$\rho = 1$~g/cm$^3$,
$\delta f \equiv |\Delta f/\Delta f_{Sauerbrey}|$,
$\Delta f_{Sauerbrey}=-h\rho\omega/2\pi m_{q}$.}
\end{figure}

\section{Liquid phase QCM applications}

Now, to complete the theory, we derive the acoustic response 
of a quartz oscillator covered by a
soft bilayer when the system is immersed in a bulk water solution.
In general, the thickness and the mechanical properties of the two layers
can be different. Such sandwich structures
are typical for  biological QCM experiments \cite{12,19,25} 
where the first layer can be a host polymer, an amphiphilic
 self-assembled monolayer
or a Langmuir-Blodgett film serving as a matrix for the second, tethered 
polymer- or adsorbed protein layer or lipid membrane. 
Water is an important component maintaining the native structure of biological
molecules and biomembranes.  

The no-slip acoustic impedance of a ``three-layer'' system 
is given by the formula
\begin{equation}
\tilde{Z}_{b} =  \frac{\xi_{1} \mu^{\ast}_{1}}{i\omega}\frac{1-\beta 
e^{2\xi_1 h_1}}{1+\beta e^{2\xi_1 h_1}} \, ,
\label{eq.fortyone}
\end{equation}
where
\begin{equation}
\beta = \frac{ \frac{\xi_{1} \mu^{\ast}_{1}}{i\omega}
\left( 1+ le^{2\xi_2 \Delta h_1}\right) -
 \frac{\xi_{2} \mu^{\ast}_{2}}{i\omega}
\left( 1 - le^{2\xi_2 \Delta h_1}\right)}
{ \frac{\xi_{1} \mu^{\ast}_{1}}{i\omega}\left( 1+ le^{2\xi_2 \Delta h_1}
\right)+ 
 \frac{\xi_{2} \mu^{\ast}_{2}}{i\omega}\left( 1 - le^{2\xi_2 
\Delta h_1}\right)}\, ,
\label{eq.fortytwo}
\nonumber
\end{equation}
\begin{equation}
l = \frac{ \frac{\xi_{2} \mu^{\ast}_{2}}{i\omega} + 1}
 {\frac{\xi_{2} \mu^{\ast}_{2}}{i\omega}- \frac{\xi_{3}
 \eta_{3}}{i\omega}}\, , \;
\Delta h_1 = h_2 - h_1, \;
\label{eq.fortythree}
\end{equation}
and
\begin{equation}
\xi_{1,2} = \sqrt{-\frac{\rho_{1,2}\omega^2}{\mu^{\ast}_{1,2}}} =
\alpha_{1,2} + ik_{1,2},~ ~ \xi_{3} = 
(1+i)\sqrt{\frac{\rho_{3}\omega}{2\eta_{3}}} \, .
\label{eq.fortyfour}
\end{equation}
Here $h_{1},\rho_{1}$ is the thickness and density of the first layer while 
$\Delta h_{2},\rho_{2}$ refer to
the second layer, $\eta_{3}$ and $\rho_{3}$ denotes the viscosity 
and density of the water solution, $\mu^{\ast}_{1,2}$ , $\alpha_{1,2}$
 and $k_{1,2}$ are given by 
Eqs.~(\ref{eq.twentythree}), (\ref{eq.twentyfour}),
(\ref{eq.twentyseven}), and (\ref{eq.twentyeight}) for Maxwell layers or
by  Eqs.~(\ref{eq.thirtythree}) - (\ref{eq.thirtyseven}) for Voight/Kelvin
layers, respectively. The acoustic response $\Delta f$ and 
$\Delta D$ can be calculated using the general results (\ref{eq.eight}) and
(\ref{eq.nine}) together with (\ref{eq.four}) and (\ref{eq.five}), where 
$\tilde{Z}_{b}$ has to be substituted for $\tilde{Z}$.

In the next section we analyse formulae (\ref{eq.fortyone}) - 
(\ref{eq.fortyfour}) in the limiting
case  of a thin viscoelastic overlayer rigidly attached to the quartz
plate and loaded by water bulk solution on top.

\section{ Apparent ``disappearance'' of the equivalent film mass
in liquid QCM measurements.}

As shown in Section III, the viscosity and elasticity of the tested
 material affect the measured equivalent mass $m_{eq}$ of the film, 
which therefore differs from the ``true'' film mass $m_{f}$.
When the system operates in a solution, acoustic shear waves 
penetrate through the thin overlayer and interact with the bulk viscous 
medium on its top. Let us analyse how the viscosity of the solution (index
``2'') changes the dissipation and resonance frequency shift of the QCM 
covered by a viscous/viscoelastic film (index ``1'') 
and how it influences its equivalent mass.

First, we restrict our attention to the special case of a
 pure viscous overlayer. By a series expansion valid for the
thin film limit $h/\delta \ll 1$, we find  that the shift 
in resonance frequency when the QCM oscillates in bulk water
is a function of the mechanical properties of the overlayer, $\rho_1$ and
$\eta_1$,  which enter to linear order in the (small) film thickness:
\begin{equation} 
\Delta \tilde{f} \approx -\frac{1}{2\pi m_{q}}
\Biggl\{\frac{\eta_2}{\delta_2}
+h_1\rho_1
\omega -2\Biggl(\frac{\eta_{2}}{\delta_{2}}\Biggr)^{2}\frac{h_1}
{\eta_{1}}\Biggr\} \, .
\label{eq.fortyfive}
\end{equation}
At the same time, the overlayer contribution to the dissipation factor 
is very small and appears only to second order in $h/\delta \ll 1$,
\begin{equation} 
\Delta \tilde{D}  \approx \Delta D_0 
\Biggl\{ 1-\frac{2\pi h_1^2 \rho_1 f}{\eta_1}
\Biggl( \frac{\eta_2\rho_2}{\eta_1\rho_1}- 1\Biggr)\Biggr\} \, ,
\label{eq.fortysix}
\end{equation}
where 
\begin{equation} 
\Delta D_0 = 2f_0\Biggl( \frac{\eta_2\rho_2}
{\pi f\rho_q C_{66}}\Biggr)^{1/2}.
\label{eq.fortyseven}
\end{equation}
is Stockbridge's result for the dissipation of a QCM in a purely viscous
bulk liquid.
We conclude from (\ref{eq.fortyfive}) that the equivalent film mass of a
purely viscous layer, 
$m \approx m_{f}(1 - (\eta_{2}\rho_{2})/(\eta_{1}\rho_{1})$,  
is obviously equal to zero if $\eta_{2}\rho_{2}= \eta_{1}\rho_{1}$.

For a QCM probing a thin layer of a Maxwell material ($h \alpha \ll 1$, 
$hk \ll1$)  in a bulk solution, we find  that the resonance frequency shift and
equivalent mass take the same form as for a purely viscous film.
However, the dissipation in a Maxwell fluid includes a term linearly dependent
on the film thickness, which appears due to the finite elasticity $\mu_1$ of 
the material:
\begin{equation} 
\Delta \tilde{D} \approx \frac{1}{\pi f m_{q}}
\Biggl\{\frac{\eta_2}{\delta_2}+
2\Biggl(\frac{\eta_2}{\rho_2}\Biggr)^2 \frac{h_{1}\omega}{\mu_{1}}\Biggr\} \, .
\label{eq.fortyeight}
\end{equation}

Finally,  we obtain in a similar way the response of a QCM covered by a thin
Voight/Kelvin overlayer when the system operates in a liquid:
\begin{equation} 
\Delta \tilde{f} \approx -\frac{1}{2\pi m_{q}}
\Biggl\{\frac{\eta_2}{\delta_2}
+h_1\rho_1
\omega -2\Biggl(\frac{\eta_{2}}{\delta_{2}}\Biggr)^{2}\frac{h_1\omega^2
\eta_{1}}{\mu_{1}^2+\omega^2 \eta_{1}^2}\Biggr\}
\label{eq.fortynine}
\end{equation}

\begin{equation} 
\Delta \tilde{D} \approx \frac{1}{\pi f m_{q}}
\Biggl\{\frac{\eta_2}{\delta_2}+
2\Biggl(\frac{\eta_2}{\delta_2}\Biggr)^2 \frac{h_{1}\omega\mu_{1}}
{\mu_{1}^2+\omega^2 \eta_{1}^2}\Biggr\} \, .
\label{eq.fifty}
\end{equation}

When $\eta\omega \ll \mu$,  viscoelastic solid and viscoelastic
fluid films dissipate the same amount of energy. In the opposite
case  of $\eta\omega \gg \mu$, a viscoelastic solid film will dissipate
much less energy than will a Maxwell fluid film. 

The equivalent mass of a viscoelastic solid film 
$$
m_{V} \approx m_{f}(1 - \frac{\eta_{2}\rho_{2}\omega^2\eta_{1}}{\rho_{1}(
\mu^{2}_{1}+\omega^2\eta^{2}_{1})})  
$$
equals zero when $\eta =\eta_{0}$, where
$$
\eta_{0\pm} = \frac{\eta_{2}\rho_{2}}{2\rho_1}\Biggl\{1 \pm \sqrt{1-
\Biggl(\frac{2\mu_{1}\rho_{1}}{\omega\eta_{2}\rho_{2}}\Biggr)^{2}}\Biggr\} \, .
$$
When the viscosity of the top layer
vanishes, $\eta_2=0$, viscoelastic corrections to the equivalent mass 
appear only in higher order corrections in the film thickness 
(see sections 3B and 3C). 

A simple analysis of expressions (\ref{eq.fortynine}) and (\ref{eq.fifty})
reveals nonmonotonic  behavior of the dissipation factor
$\Delta D(\mu,\eta)$ and the resonance frequency shift $\Delta f(\mu,\eta)$.
The dissipation factor will reach a maximum when $\mu^{\ast}=\eta\omega$ 
if the viscosity is constant. In the same way, the resonance frequency
has a minimum as a function of viscosity when $\eta^{\ast}=\mu/\omega$ 
(Fig.~6). This circumstance must be taken into account in QCM  adsorption
experiments if viscoelastic parameters and the layer thickness can  vary
simultaneously during adsorption from the liquid phase \cite{18}.

It should be mentioned, that
the QCM characteristics in the case of a Maxwell fluid are monotonic functions
of film viscoelasticity, decreasing with $\mu$ and $\eta$, respectively. 

\begin{figure}
\centerline{\psfig{figure=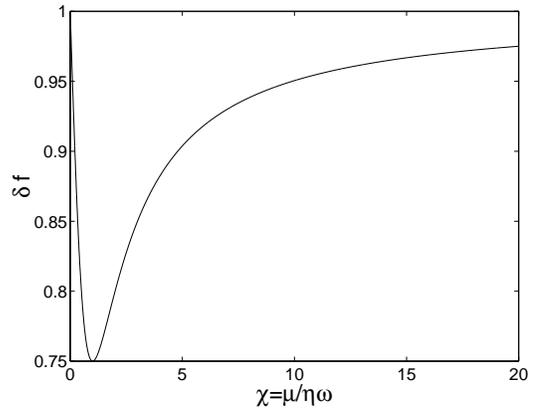,width=7cm}}
\vspace*{7mm}
\caption{Relative resonance frequency  shift  $\delta f$
due to a thin viscoelastic solid overlayer as a function of 
its viscoelastic ratio $\chi=\mu/\eta\omega$. QCM operates in
 a bulk aqueous solution; $\delta f = 
|(\Delta f - \Delta f_{0})/\Delta f_{Sauerbrey}|$,
$f_{0}$ is the resonance frequency shift in a bulk liquid without overlayer;
$\Delta f_{0}= - \eta/2\pi m_{q}\delta$, 
$\Delta f_{Sauerbrey}=-h\rho\omega/2\pi m_{q}$.}
\end{figure}

\section{Discussion: soft matter and biosensor applications of QCM}

One of the current challenges is to apply the QCM technique to probing 
biological materials in solution: polymers and complex fluids, which are
dominated by viscous effects. Typically adsorbed from a liquid phase, they
form soft biological interfaces of multilayer architecture at the 
solid substrate. Recently it has been reported that mechanical properties
of (bio)polymer and amphiphilic self-assembled films, adsorbed films of
proteins, enzymes, membrane microspheres (vesicles) and even living cells can
be investigated by using quartz crystal oscillators \cite{12,13,18,18,25,26}.
It seems crucial that an adequate physical model is chosen for a
quantitive interpretation of such QCM measurements to be possible. 
In addition, biosensor applications of the quartz crystal microbalance
(QCM) technique are faces with the need to control the interfacial friction, 
when the probed biopolymer material can slip relative 
to the quartz surface during its oscillatory motion.
 The slip of the adjacent layer, affecting the resonance frequency
shift $\Delta f$ and dissipation factor $\Delta D$ of the oscillator arises 
from the weak coupling between film and substrate.

The limit of strong coupling of the layer to the solid surface 
corresponds to the widely used ``no-slip" boundary condition. 
Experimentally, this boundary condition applies to thin solid films 
evaporated onto the quartz surface.
Among biomaterials, the no-slip conditions can easily be realized
for Langmuir-Blodgett films, which can be extremely strongly coupled to 
the solid substrate. 

 The usage of QCM for biological and biomedical purposes  brings out
the problem of accounting for the complex rheology of the biological interface 
formed by the adsorbed biomolecular films. 
In some special cases or for particular type of shear deformations,
 the  linear viscoelasticity of polymers 
can be treated within either the Maxwell or the Voight/Kelvin schemes.
The results of the Voight/Kelvin model can readily be applied
to quartz crystal acoustic measurements of adsorbed proteins,
which conserve their shape during adsorption and do not flow under
shear deformation as well as to polymer films far from the glass
transition region. The Maxwell fluid scheme is appropriate for 
polymer solutions that are pure liquids (at low shear rates) \cite{23}
and polymers in the amorphous state and in the vicinity of the polymer 
liquid-glass transition.
Viscoelastic materials with more complex rheology (e.g., cells, 
membranes, liquid crystal polymers etc.)  can be described by 
a combination of these two basic 
viscoelastic schemes. Two such schemes are illustrated 
in Figs.~3c and 3d. They can be analysed
in a similar as we have done here way by substitution of the corresponding
expressions for the complex shear modulus $\mu^{\ast}$ in the general results
(\ref{eq.twentysix}) - (\ref{eq.twentyeight}) and then in 
(\ref{eq.eight}) - (\ref{eq.nine}).  For
instance, the complex shear modulus for the material described by scheme of
Fig.~3d, is given by the expression 
\begin{equation}
\mu^{\ast -1} = \frac{\mu_{1}+i\omega\eta_{1}}
{i\omega\eta_{1}\mu_{1}} + \frac{1}{\mu_{2}+i\omega\eta_{2}}.
\end{equation}
Such a model well describes, e.g., the viscoelasticity  of the lyotropic
lamellar phase (the so-called ``onion phase'') which organizes itself into
multilamellar vesicles which are closed-packed and fill up
the space \cite{24}.

It is convenient to describe the frequency behavior of a complex viscous medium
exhibiting a spectrum of relaxation times in 
terms of  the three-parameter model 
 $\eta^{\ast}=\eta/(1+i\omega\tau)^b$ of the complex viscosity
suggested by Reed, Kanazawa and Kaufman \cite{17}.
In this model, $\eta$ is the viscosity of the liquid,
 $b$ is a characteristic exponent
related to the distribution of relaxation times $\tau$.
In the same manner, we can introduce in our model a complex shear modulus 
of the form
$$
\mu^{\ast}=\mu/(1-i/\omega\tau)^{b}
$$ 
which is a generalization for the case of a relaxation time 
distribution characterized by the exponent $b$. In particular, the Maxwell
fluid corresponds to the value $b=1$.

In summary, 
recent results show that in cases where the adsorbed material forms a
soft interface to the quartz crystal microbalance (QCM),  both internal 
and interfacial friction mechanisms may contribute to energy
dissipation. The combined action of these mechanisms may cause the variation 
of the shift $\Delta D$ in the dissipation factor to be a nonmonotonic function
of material parameters, which can be varied in an experiment. Depending on the
system, one or the other type of friction may dominate. We have suggested here
a dynamic ``slip/no-slip''  test that can be applied to QCM measurements
with thin adsorbed films. The test is based on the frequency dependence of a
peak in the dissipation factor.  We have shown that the peak value of
$\Delta D$ caused by sliding friction is frequency independent while the
viscoelasic peak is strongly dependent on frequency.

Besides being able to understand the role of interfacial friction, an important
problem for the application of the QCM technique to biosensoring remains.
How can the device recognize what type of material is being probed except by
detection  of an ``anonymous'' deposited mass? In this article we
suggest that analysing the frequency behavior of QCM characteristics as one
of possible test what  material is overlayered the quartz surface. 

Another way is a direct biochemical modification of QCM
surface for selective binding of tested biological molecules.
Recently, an immunosensing system based on antigen-coated QCM
has been reported elsewhere \cite{27}. It has been shown that 
QCM  results were in agreement with those of enzyme linked 
immunosorbent assay (ELISA). Perhaps the successful road to QCM usage
as a biosensor device, is to combine both of dynamic measurements 
and immunology treatment directions. 

Presented here rigorous formulae 
for experimentally measured $\Delta f$ and $\Delta D$ values,
make it possible to investigate viscoelastic contribution in a wide range of
biomolecular materials and can be useful for the correct interpretation
of biologically oriented QCM experiments.

\section{ Acknowledgements.}

This work was supported by the Royal Swedish Academy of Sciences 
(KVA) and the Swedish TFR. One of us (MVV) thanks  
Dr.~Bo~Persson for fruitful discussions of the theory of sliding friction and
for making Ref.~\cite{9} avaliable before publication.
We are indebted to Dr.~M.~Rodahl for helpful
cooperation and  many discussions about QCM experiments on complex fluids and
biomolecular layers and to Drs.~A.~Krozer, C.~Fredriksson, F.~H{\"o\"o}k
and Ms. K. Glasm{\"a}star for stimulating discussions about QCM 
biosensor applications.

\end{document}